# Machine Learning Enabled Graph Analysis of Particulate Composites: Application to Solid-state Battery Cathodes

Zebin Li[1], Shimao Deng[2], Yijin Liu[2*], Jia-Mian Hu[1*]

[1]*Department of Materials Science and Engineering, University of Wisconsin-Madison, Madison, WI 53706, USA*

[2]*Walker Department of Mechanical Engineering, University of Texas at Austin, Austin, TX 78712, USA*

## Abstract

Particulate composites underpin many solid-state chemical and electrochemical systems, where microstructural features such as multiphase boundaries and inter-particle connections strongly influence system performance. Advances in X-ray microscopy enable capturing large-scale, multimodal images of these complex microstructures with an unprecedentedly high throughput. However, harnessing these datasets to discover new physical insights and guide microstructure optimization remains a major challenge. Here, we develop a machine learning (ML) enabled framework that enables automated transformation of experimental multimodal X-ray images of multiphase particulate composites into scalable, topology-aware graphs for extracting physical insights and establishing local microstructure-property relationships at both the particle and network level. Using the multiphase particulate cathode of solid-state lithium batteries as an example, our ML-enabled graph analysis corroborates the critical role of triple phase junctions and concurrent ion/electron conduction channels in realizing desirable local electrochemical activity. Our work establishes graph-based microstructure representation as a powerful paradigm for bridging multimodal experimental imaging and functional understanding, and facilitating microstructure-aware data-driven materials design in a broad range of particulate composites.

*E-mails: liuyijin@utexas.edu (Y.L.) or jhu238@wisc.edu (J.-M.H.)

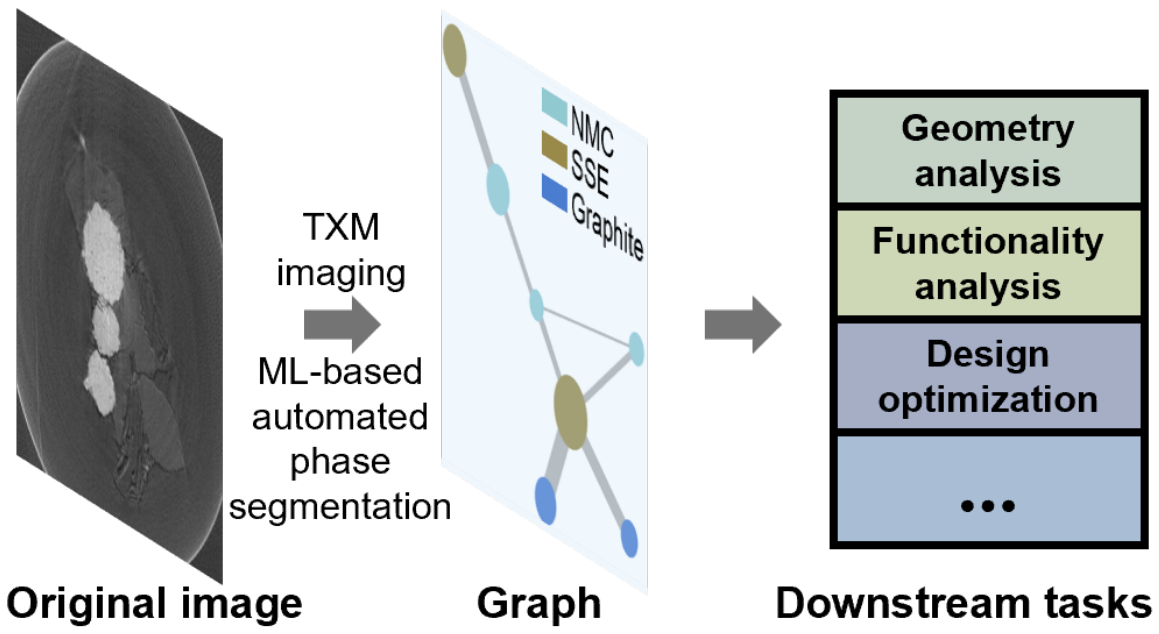

Particulate composites are widely used in solid-state chemical and electrochemical systems, including solid-state batteries (SSBs) [1], solid oxide fuel cells (SOFCs) [2], catalysts for fluid catalytic cracking (FCC) in the petroleum industry [3], and advanced sensors [4]. The microstructure of these particulate composites, notably the phase boundaries and inter-particle connections, can critically affect their properties and the system performances. For example, in the process of refining petroleum through FCC, the catalytic active sites are the ultimate drivers that chemically convert the feedstock into the targeted chemicals [3]. In SOFCs, the three-phase (i.e., cathode, electrolyte, and air) boundaries are the active sites for oxygen reduction reaction [5]. Similarly, in SSBs, the active materials, ionic conductors, and electrical conductors need to work together seamlessly to enable the on-demand energy storage and release.

Although the critical role of the microstructure is well recognized, its impact on the properties of particulate composites remains challenging to quantify. Recently, Advances in electron and X-ray microscopy tomography enable the acquisition of large-volume, multimodal images for microstructures of particulate composites[6]. To connect these high-dimensional microstructure datasets to the properties and performance, and in future, the parameters of synthesis, processing, and manufacturing, it is necessary to obtain a lower-dimensional representation that captures the most salient features of a given microstructure. Specifically, for representing the microstructures of multiphase particulate composites, existing methods, which are based on the image pixels/voxels [7-12] or statistical correlations [13-17], would not suffice for two reasons. First, it would be computationally expensive to apply them to microstructures of practical size that contain billions of voxels. Second, the connectivity and topological characteristics of the particles and other constituent phases, which critically determine the materials properties and system performance, cannot be encoded into the low-dimensional representation via these methods.

To address this challenge, here we use graphs to represent the microstructures of particulate composites, which has three main advantages. First, since the individual particles/phases almost always occupy more than one pixel/voxel in the raw data, graph-based representation would be computationally more efficient than image-based representation and permits incorporating thousands of particles with minimal memory cost [18]. Second, by representing individual particles/phases as nodes and their interfaces as edges, the connectivity and topology of the particles/phases are naturally retained. Third, by storing both the structural and functional features of individual particles/phases as well as the interfaces into the node and edge feature vectors, respectively, graph would facilitate multiscale, multimodal information fusion and the development of local microstructure-property relationships. Thus far, graphs have been extensively used to represent polycrystalline microstructures, which has in turn enabled the application of graph neural network (GNN) to link the low-dimensional embedding of the polycrystal graph to properties [18-23]. However, to our knowledge, graphs have not yet been applied to represent the microstructures of multiphase particulate composites. This is in part due to the challenge of achieving an accurate and automated segmentation of the individual phases from the experimentally measured raw microstructure images, which can now thankfully be addressed by leveraging advanced machine learning (ML)-powered computer vision models [24-29].

As an example, here we demonstrate the use of graphs to represent the experimentally measured microstructure images of the composite cathode of SSBs, which is comprised of mixed active particles ($LiNi_xMn_yCo_zO_2$, $x + y + z = 1$, donated as NMC), solid-state electrolytes (SSEs), and conductive carbon (graphite) that we show can be automatedly segmented by ML models. We then discuss how such graph-based representation can be used to facilitate the study of the microstructure-property relationships at both the particle and the network level, and to evaluate the alignment of microstructure design rules. To that end, we first acquire high-resolution, chemically resolved imaging of SSB cathodes using full-field X-ray imaging, and then develop a workflow that capitalizes on ML-based phase segmentation to automatically convert the original X-ray images to graphs. Notably, both the morphological features and the local electrochemical properties of NMC particles (i.e., Ni oxidation state, represented by the Ni K-edge energy[30-32]) are incorporated into the feature vectors of the nodes/edges of the graphs. Building on such node-/edge-level correlations, we establish an understanding of the local microstructure-property relationship in such a complex mesoscopic system. Last but not the least, we demonstrate that graph enables a GNN-based property prediction at the node level, representing a critical step for the realization of microstructure-aware inverse design of processing and synthesis parameters that yield the desirable material properties.

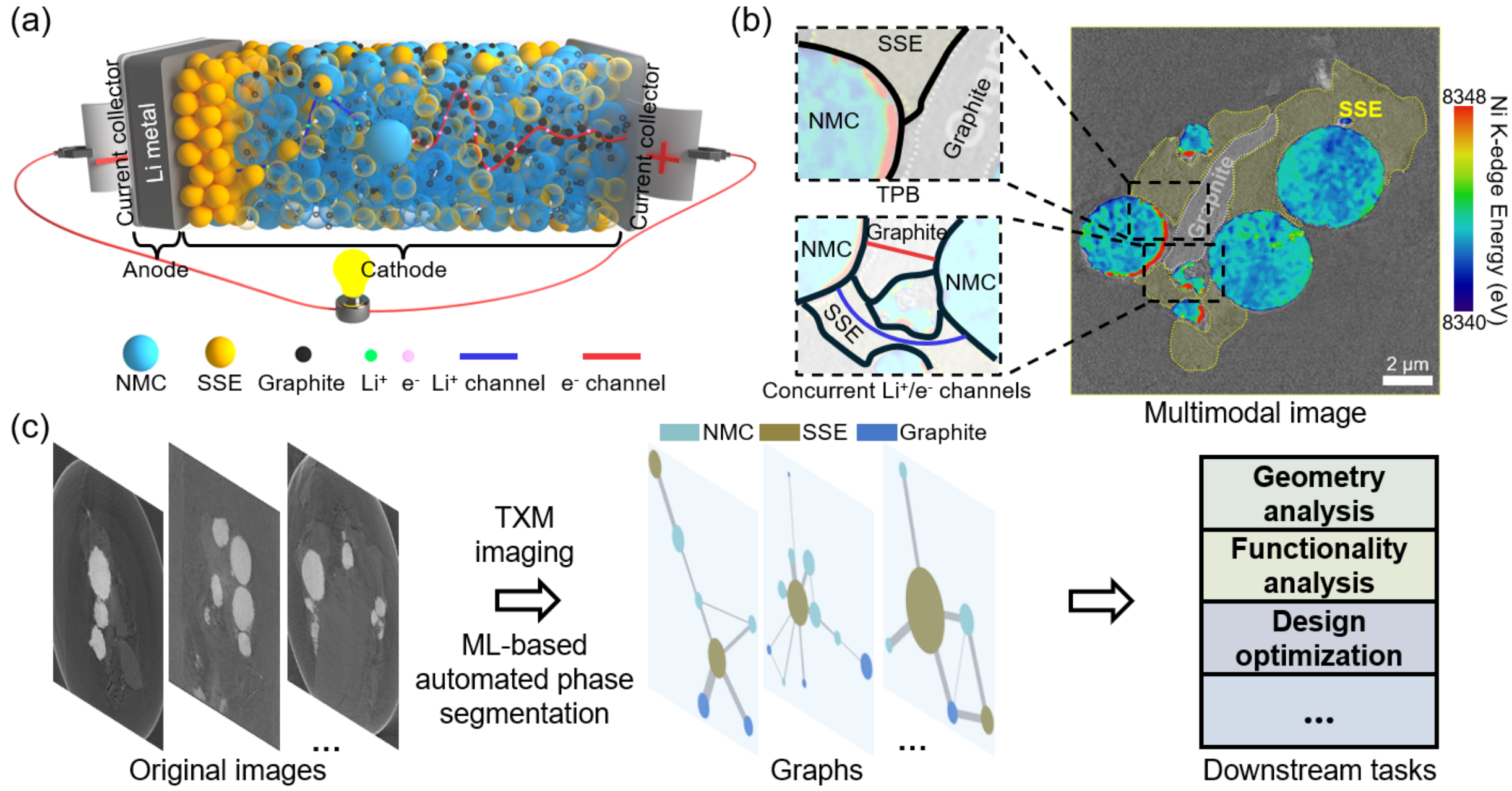

**Figure 1. Graph representation and analyses of multiphase particulate composites, using the composite cathode of the solid-state battery (SSB) as an example. (a)** Illustration of an SSB. **(b)** Example of a multimodal image integrated from a 2D X-ray image with human-expert-annotated phases and the corresponding local electrochemical states (i.e., Ni oxidation states, represented by the Ni K-edge energy) of NMC particles obtained by TXM imaging. Zoom-in figures demonstrate the two critical microstructure features, i.e., triple phase boundaries (TPBs) and concurrent $Li^+$/$e^-$ channels via the connecting SSE/graphite. **(c)** The ML-enabled workflow for automatedly converting raw X-ray images into microstructure graphs for facilitating and enabling various downstream tasks.

The demonstration of the SSB system is shown in **Fig. 1a**, where the cathode is comprised of a mixture of NMC particles, SSEs, and graphite. During discharging, $Li^+$ and electrons migrate through the networks of SSEs and graphite, respectively, both ultimately reaching the NMC particles for Li intercalation. Such an intercalation reaction would be most efficient at the NMC-SSE-graphite triple phase boundaries (TPBs), where the concentrations of $Li^+$ and $e^-$ are simultaneously high. An equally important microstructure feature is the concurrent $Li^+$/$e^-$ channels between NMC particles. For example, NMC-SSE-NMC and NMC-graphite-NMC connections can provide channels for the migration of $Li^+$ and $e^-$, respectively. To analyze these local microstructure features and their correlations with local properties, we apply energy-resolved transmission X-ray microscopy (TXM), which enables the simultaneous spatial mapping of phase morphology and the electrochemical states of NMC particles, resulting in the multimodal imaging (**Fig. 1b**). The zoom-in figures of **Fig. 1b** highlight the TPBs (top panel) and the concurrent $Li^+$/$e^-$ channels (bottom panel). The original X-ray image and the corresponding TXM image are shown in Supporting **Fig. S1**. **Figure 1c** shows the workflow of converting raw X-ray images into graphs and the downstream analyses that relate local geometrical and topological features to local electrochemical states.

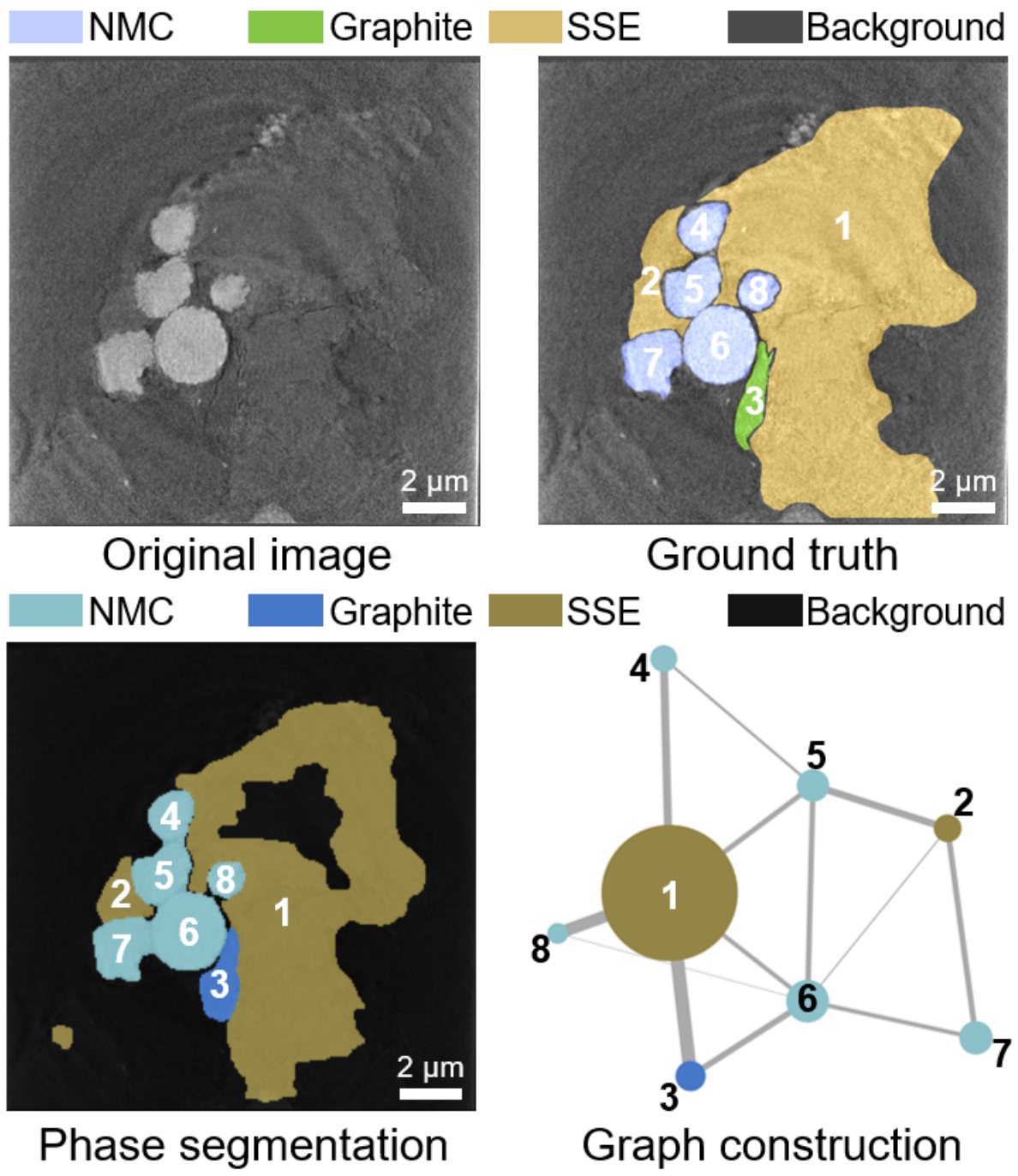

**Figure 2. Example results of graph construction enabled by ML-based automated phase segmentation.** In the constructed graphs, the node sizes and the edge thicknesses represent the sizes and the interface areas of the corresponding objects and connections, respectively.

**Figure 2** shows an example (from the testing dataset) of the original X-ray image of SSB cathodes, the phase segmentation results from a human expert (ground truth) and a customized U-Net (see Supporting Information), and the constructed graph. More examples are shown in Supporting **Fig. S2**. Moreover, based on the instance segmentation map (Supporting **Fig. S3**), we assign identical numbers to each object in the segmented phase map and the corresponding node in the graph. Because its end nodes are explicitly labelled, each edge of a graph, corresponding to the interface of two neighboring objects, can also be uniquely identified. Together, the microstructure graph we constructed allows for compiling physical features for each node and edge and digitalize the neighboring relationships of these individual objects in the form of an adjacency matrix, which are necessary for the application of GNN to link either microstructure graphs[20-23] or atomic-scale structure graphs[33-37] to properties.

Despite the challenges arising due to the similar pixel intensity distributions of monolithic phases in the original X-ray images (**Fig. S1b**), the U-Net based phase segmentation is largely successful, especially for NMC particles. The automatedly constructed graph (bottom right panel of **Fig. 2**) mostly captures the real geometric connections, where the neighboring objects are connected by edges. The edge thickness is proportional to the interface area, which is quantified by the number of overlapped pixels. The node size is proportional to the pixel count of each object.

Recent works have demonstrated that the design of microstructure morphology is key to achieve efficient charge-discharge at the particle level [38]. The spatial arrangements of NMC-graphite and NMC-SSE contracts, as well as the resulting TPBs, are pivotal for providing balanced fluxes of $Li^+$ and $e^-$ ($J^{Li} \approx J^{e}$) for active cathode materials. Therefore, the optimization of NMC particle size is a nontrivial task. Smaller particles exhibit shorter diffusion length within the NMC particle and are more favorable for fast charging applications. On the other hand, large particles have a higher probability of accommodating multiple TPBs on their surface, integrating the particle into well-configurated conductive networks for both $Li^+$ and $e^-$. Despite tremendous optimization efforts, the exiting manufacturing method does not have a full control of the electrode micromorphology. Therefore, the high-resolution and large field-of-view microscopic characterization is critically needed to inform the improvement of the manufacturing protocol.

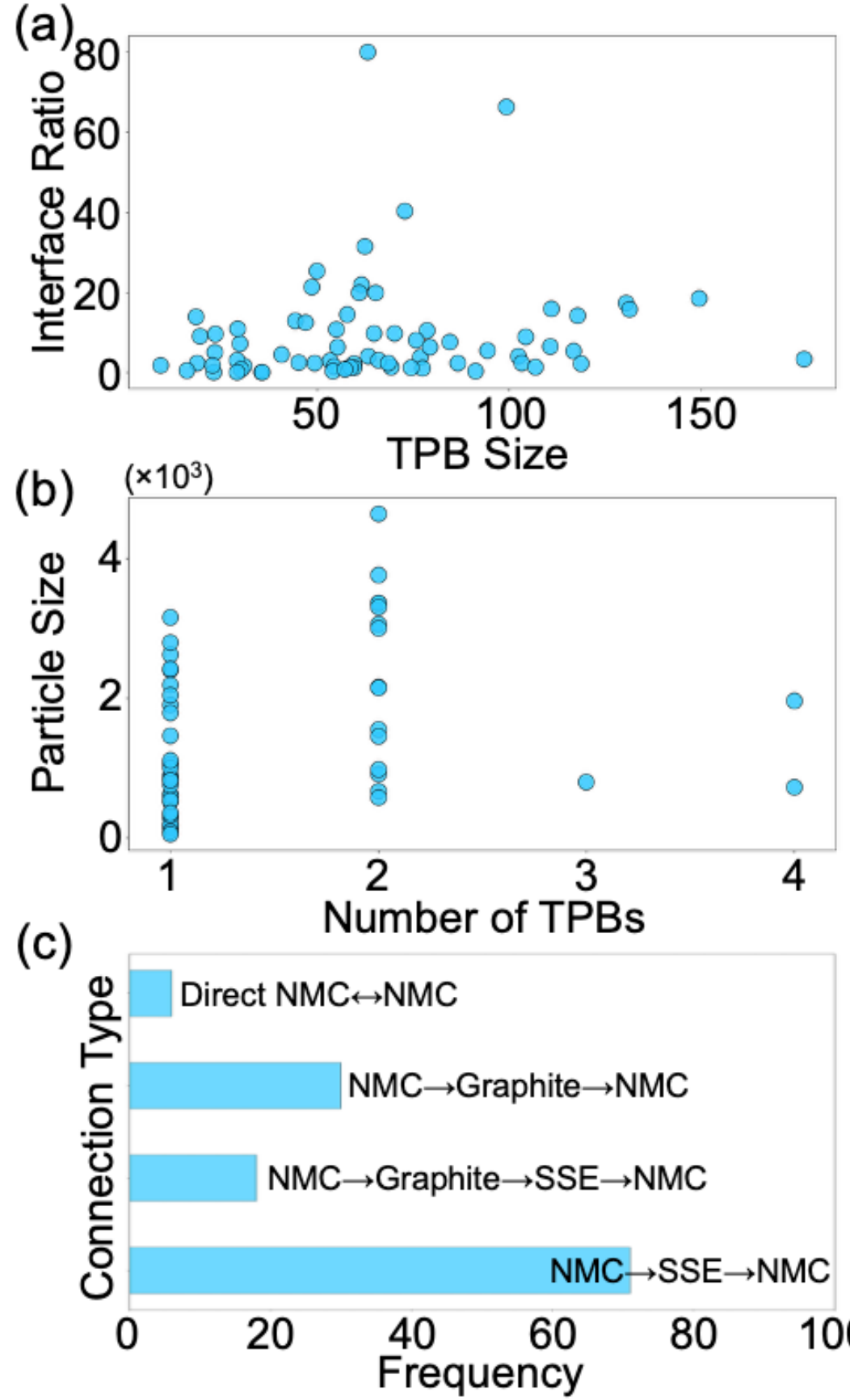

**Figure 3. Microstructure morphology analysis. (a)** Distribution of TPBs with respect to the TPB size (i.e., the perimeter of NMC-SSE-graphite triangle), and the interface area ratio (i.e., the edge weights for NMC-SSE over NMC-graphite). **(b)** Distribution of NMC particles with respect to the number of associated TPBs, and the particle size (quantified by pixel count). **(c)** Barplot of NMC particle inter-particle connections (i.e., the channel to the nearest NMC particle).

In addition to the experimental characterization efforts, high-throughput quantification of the imaging data is also indispensable. Although one can evaluate whether the cathode microstructure of SSBs meets these design targets by directly analyzing the phase segmented microstructure image, however, it is often a tedious and labor-intensive process. Instead, performing such evaluation on microstructure graphs will enable automated and high-throughput statistical analyses. For demonstration, we apply our ML-enabled graph construction to 73 new original X-ray microstructure images, yielding automated construction of 73 microstructure graphs. **Figure 3a** shows the distributions of TPB size (precisely speaking, the perimeter of TPB) with respect to the ratio of the NMC-SSE interface area to the NMC-graphite interface area, which affects $J^{Li}/J^{e}$ by modulating their respective interfacial resistance. An ideal microstructural configuration would offer a large TPB size and a balanced and consistent value of interfacial area ratio, thus stabilized and optimized $J^{Li}$ and $J^{e}$ for every single cathode particle. The data in **Fig. 3a**, however, demonstrates that most of TPBs concentrate in the left part and have ratios that are significantly scattered. **Figure 3b** shows the distribution of the NMC particle size (pixel count) with respect to the number of TPBs each particle is connected to (only those NMCs associated with TPBs are counted). The data shows that most NMC particles are only involved in one or two TPBs, which is less than ideal. Furthermore, it is desirable to establish concurrent $Li^{+}/e^{-}$ channels by connecting two NMC particles via both the SSE and graphite (see the bottom inset in **Fig. 1(b)**). Yet, as shown in **Fig. 3c**, most of NMC particles are only connected to SSE, with limited direct contact with graphite (hence the $J^{Li}$ and $J^{e}$ are imbalanced). These morphology analyses indicate that there is plenty of room for improving the performance of SSBs by microstructure engineering.

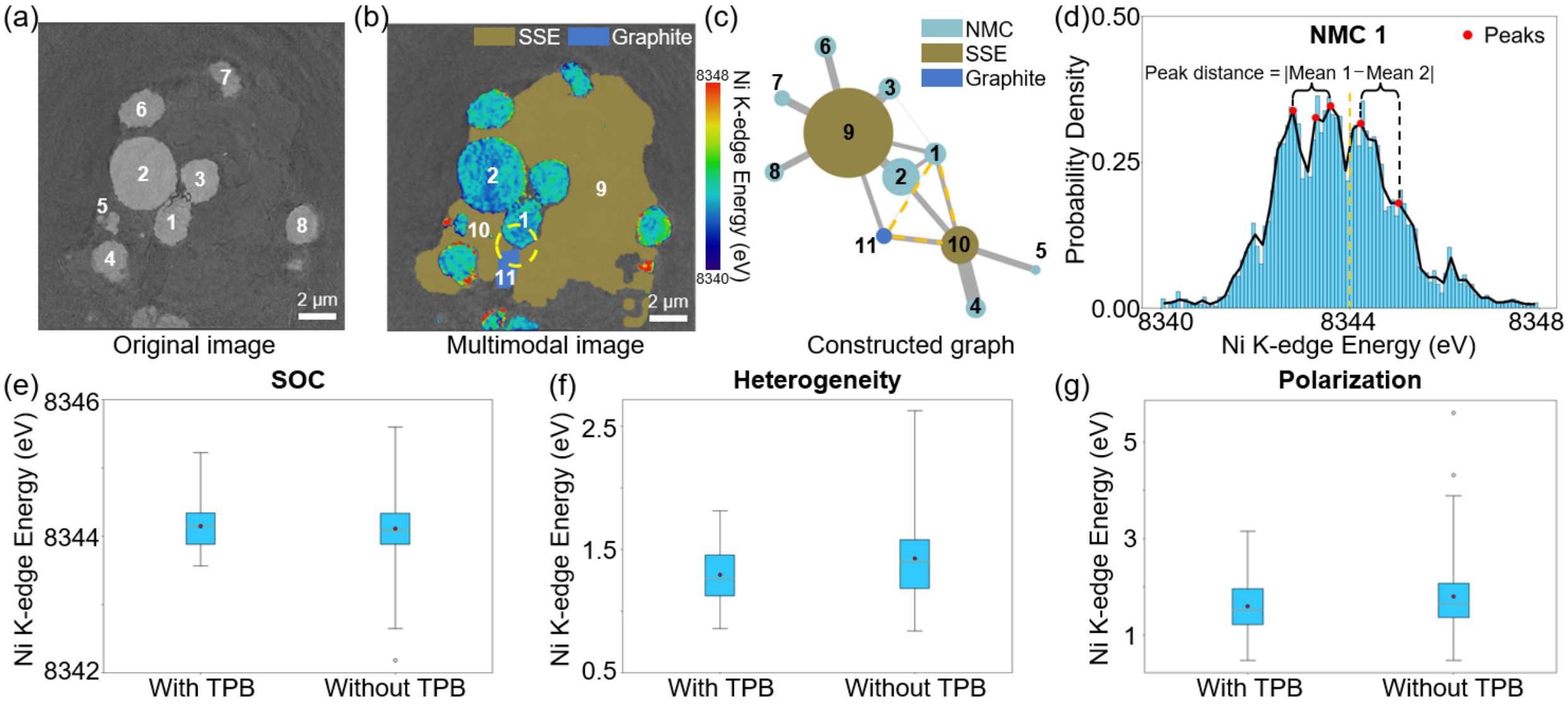

**Figure 4. Microstructure-property relationship at the node level.** Example of **(a)** the original image, **(b)** the corresponding multimodal image, and **(c)** the constructed graph. **(d)** The distribution of the electrochemical state (i.e., Ni oxidation state represented by Ni K-edge energy) for #1 NMC in **Fig. 4a**. The distributions of the remaining NMCs are shown in Supporting **Fig. S4**. The electrochemical states are described by their distribution means, standard deviations, and peak distances, which are the measurements of intra-particle SOC, heterogeneity, and polarization, respectively. Detected peaks (≥40% of the maximum height) are marked in red dots. Peak distance is calculated as the distance between the mean locations of peaks smaller and larger than 8344 eV.

Boxplots of NMC particles with and without TPBs in terms of **(e)** the SOC, **(f)** the heterogeneity, and **(g)** the polarization. The short horizontal orange line represents the median and the small dot shows the mean value. A lower value represents a higher degree of completion for the electrochemical reaction which is desirable.

We demonstrate the use of microstructure graphs to examine the correlation between the local microstructure morphologies and the local electrochemical states of individual NMC particles. The local electrochemical states are obtained from energy-resolved three-dimensional full-field TXM (see Supporting Information). **Figure 4a-c** present the multimodal imaging example that reflects both the local electrochemical states of NMC particles and the ML-segmented phases, as well as the corresponding graph. It is worth noting that the cathode samples used in this work are discharged, at which a lower Ni oxidation level (represented by lower values of the Ni K-edge energy) indicates a higher degree of completion for the electrochemical reaction. We made two main observations from **Figs. 4a-c**. First, the NMC particle (i.e., particle #1 in **Fig. 4b**) involved in the TPB (labeled by the yellow dashed circle in **Fig. 4b** and the dashed triangle in **Fig. 4c**) has lower Ni oxidation states than the other NMC particles. Second, NMC particle #2 also exhibits a low Ni oxidation state due to its electronic connection with NMC particle #1, given that NMC is also an electron conductor [39, 40]. In addition, $Li^+$ can access the surface of particle #2 through the surrounding SSE. These findings corroborate the significant impact of the inter-particle connectivity on the electrochemical states of NMC, which will be discussed in greater details in the next session.

**Figure 4d** shows the statistics of the local electrochemical state (represented by the Ni K-edge energy in each pixel) for the #1 NMC particle in the multimodal image, and those of the remaining NMC particles are shown in Supporting **Fig. S4**. The means, standard deviations, and peak distances in these statistical distributions, as indicated in **Fig. 4d** (see details in Supporting Information), correspond to the state of charge (SOC), heterogeneity, and polarization of the NMC particle, respectively. We note that using relative K-edge shifts to evaluate SOC variations is a widely adopted approach in the battery TXM community[30-32, 41], particularly when comparing samples measured under identical beamline configuration and energy calibration as in this study. We then investigate all the NMC particles in our dataset, where 48 out of 350 NMC particles are involved in TPBs, and examine the effect of TPBs involvement on the SOC, heterogeneity, and polarization. As shown in **Figs. 4e-g**, while the averaged SOCs of the NMC particles remain largely independent of their involvement in a TPB, the values of heterogeneity and polarization in NMC particles with TPB are lower than those without TPBs, indicating a more homogeneous and efficient electrochemical reaction. In this regard, our results indicate that the presence or absence of TPB involvement significantly influences the spatial distribution of Ni oxidation states even after significant relaxation. This finding highlights the role of TPB on reaction uniformity as an intrinsic, sustained structural constraint that persists beyond transient polarization effects. Moreover, while the local electrochemical behaviors can also be influenced by other factors including particle size, contact area, local porosity, and proximity to boundary layers, our detailed analyses show that the TPB abundance captures the dominant structural contribution among these correlated factors (see Supporting Information). These results confirm that TPBs promote a spatially more uniform charge-discharge behavior inside an NMC particle and thereby contribute to a prolonged battery lifespan. In particular, the TPB involvement is not considered as a direct, standalone cause of relaxed-state intra-particle heterogeneity, but rather as a key structural factor shaping the surface reaction distribution, which in turn contributes to the development and persistence of internal heterogeneity under solid-state conditions. Furthermore, it is worth noting that the measured Ni oxidation-state distribution reflects the residual heterogeneity after substantial relaxation, rather than the reaction nonuniformity driven by transient polarization under current flow. We believe that the persistence of this heterogeneity after relaxation suggest that significant reaction nonuniformity had developed during cycling, because relaxation is generally expected to reduce the heterogeneity of the Ni oxidation-state distribution.

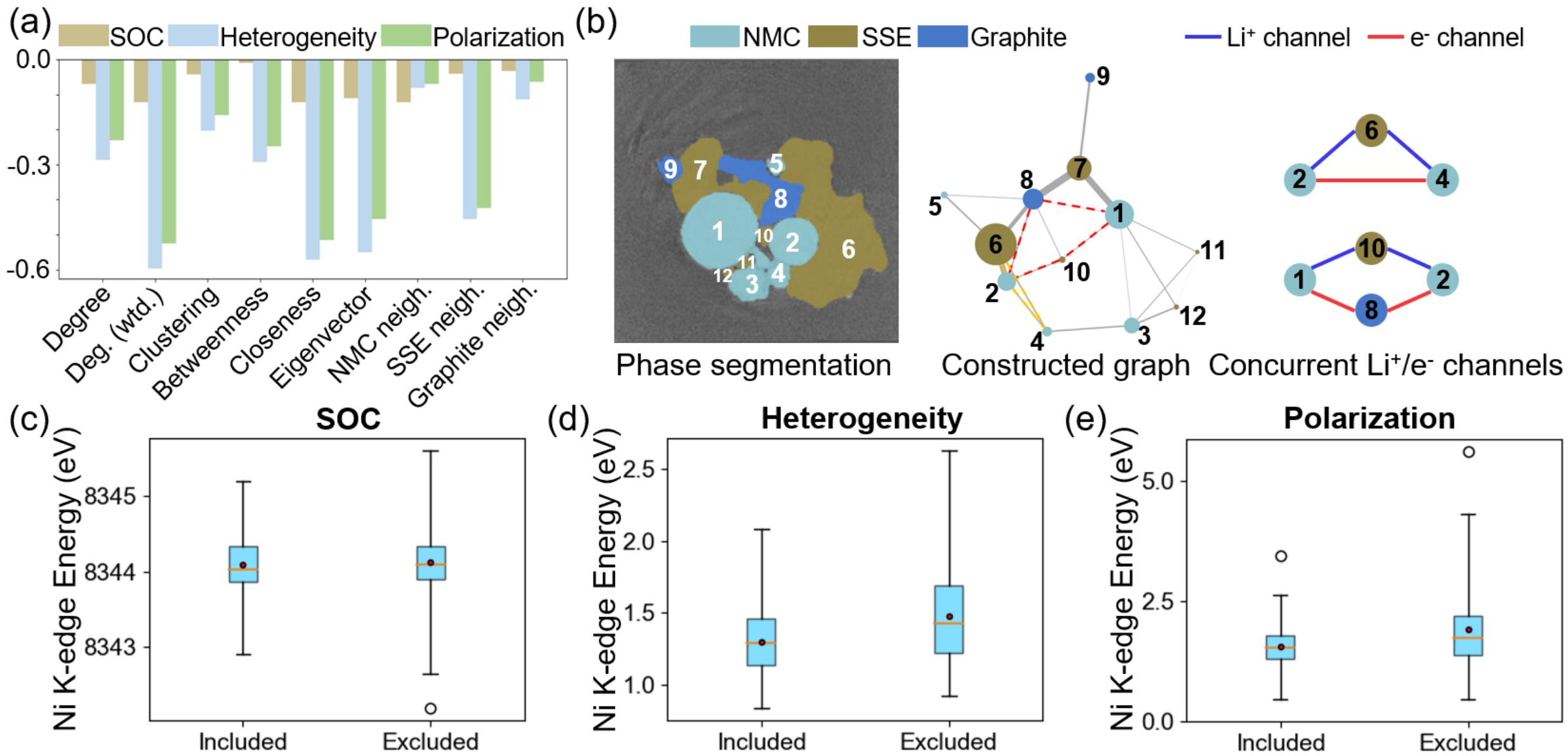

**Figure 5. Microstructure-property relationship at the network level. (a)** Pearson correlation coefficients between graph-theoretic metrics and electrochemical states (i.e., intra-particle SOC, heterogeneity, and polarization). **(b)** Demonstration of two types of inter-particle connections that entail concurrent $Li^+$/$e^-$ channels in both the segmented X-ray image and constructed graph. Boxplots of NMC particles included and excluded in the inter-particle connections in terms of **(c)** the intra-particle SOC, **(d)** the heterogeneity, and **(e)** the polarization. The short horizontal orange line represents the median and the small dot shows the mean value. A lower value represents a higher degree of completion for the electrochemical reaction which is desirable.

Microstructure graphs also allow for investigating the microstructure-property relationship at the network level, leveraging established graph analysis tools and theories. Specifically, we quantify key graph-network-level descriptors for NMC particles, including degree, centrality indicators (e.g., closeness, betweenness, and eigenvector), clustering coefficient, and then correlate them with NMC electrochemical states by computing the Pearson correlation coefficient ($r$) (see Supporting Information). A more negative $r$ indicates a stronger correlation and a lower Ni K-edge energy (i.e., a more complete electrochemical reaction). Degree quantifies the number of neighbors for a node. Centrality indicators quantify the importance and influence of a node within the overall network, where closeness reflects proximity to all other nodes, betweenness identifies nodes acting as bridges along shortest paths, and eigenvector highlights nodes connected to other influential nodes. Clustering coefficient measures the tendency of a node's neighbors to form tightly connected groups. As shown in **Fig. 5a**, the heterogeneity and polarization of NMC particles are both strongly correlated with the network descriptors ($r$ is large) and have similar correlation patterns, while the correlation between SOC and descriptors is weaker, consistent with the result in **Fig. 4e**. The high absolute $r$ values of closeness and eigenvector indicate that the NMC particles are more deeply embedded within the network, which is consistent with their large number of SSE neighbors and has strong correlation with their electrochemical states. In addition, the high weighted degree compared to degree suggests the importance of taking the edge weight (i.e., interface area) into consideration since the interface area affects the efficiency of $Li^+/e^-$ transport. The scatter plots of the top four graph-theoretic metrics with the strongest absolute $r$ with the electrochemical states, i.e., intra-particle SOC, heterogeneity, and polarization are shown in Supporting Information **Figs. S5-S7**.

The left panel of **Fig. 5b** shows a segmented microstructure image with NMC particles that have the two types of desirable inter-particle connections, which are also highlighted as the dashed lines in the constructed graph (see the middle panel of **Fig. 5b**). The right panel of **Fig. 5b** illustrates these two types of desirable connections between two NMC particles that both can provide concurrent ionic/electronic channels. To further validate this finding, we evaluate the impact of the concurrent $Li^+/e^-$ channels on the electrochemical states (i.e., SOC, homogeneity, and polarization) for a total of 350 NMC particles, among which 136 particles are connected through concurrent $Li^+/e^-$ channels. As shown in **Figs. 5c-e**, the 136 NMC particles with these concurrent channels exhibit reduced heterogeneity and polarization compared to the remaining 214, whereas their SOC shows minor differences. This result unravels the importance of building concurrent $Li^+/e^-$ channels among the NMC particles, in addition to involving them in a TPB.

Importantly, in a recent experiment[38] (led by a subset of the authors of this work), the graphite/SSE ratios are systematically varied to construct three types of composite cathode with $Li^+$-channel-deficient (LCD), $e^-$-channel-deficient (ECD), balanced $Li^+/e^-$ (BLE) configurations. Among them, the BLE configuration, which inherently corresponds to a higher abundance of effective TPBs, exhibited optimized capacity and cycling stability. In contrast, ECD electrodes showed reduced capacity and shortened lifetime, while LCD electrodes failed to operate properly due to insufficient ionic percolation. In the present study, we employ ML-enabled graph analysis to quantitatively map TPB distribution and concurrent transport channels. The results align with the performance trends reported in ref. 38 and support the proposed interpretation that the abundance of TPBs and concurrent $Li^+/e^-$ channels could influence the performance of SSBs.

Automated construction of microstructure graphs from original multimodal microstructure images also enables the application of GNN as a surrogate model to predict the local microstructure-property relationship. **Figure 6** demonstrates the prediction of the electrochemical states in each NMC particle via GNN (see Supporting Information). The prediction performances for the intra-particle heterogeneity and polarization are better than SOC, which are consistent with the previous graph-based analyses showing no significant association between SOC and the TPB (**Fig. 4e**) or the concurrent $Li^+/e^-$ channels (**Fig. 5c**). The overall prediction results of these three descriptors indicate limited predictive capability, which can be attributed to several factors. These factors may include the small dataset size, the noise in experimental data, the potential loss of information resulting from the use of two-dimensional (2D) slices instead of three-dimensional (3D) images (specifically, NMC particles that appear to lack TPBs in a 2D slice may in fact possess TPBs in a 3D image), and the relatively small representative microstructure volume (a battery cell contains thousands of NMC particles which significant exceeds the volume sampled in the present experiment). Although the roles of these factors can only be quantified when larger-scale 3D microstructure data with improved phase-contrast become available (acquiring those data would require more advanced tomography techniques), the demonstration in the present work successfully shows the potential of leveraging graph representation and graph learning to directly perform node-level prediction tasks on the experimental data, setting the foundation for the graph-based microstructure inverse design.

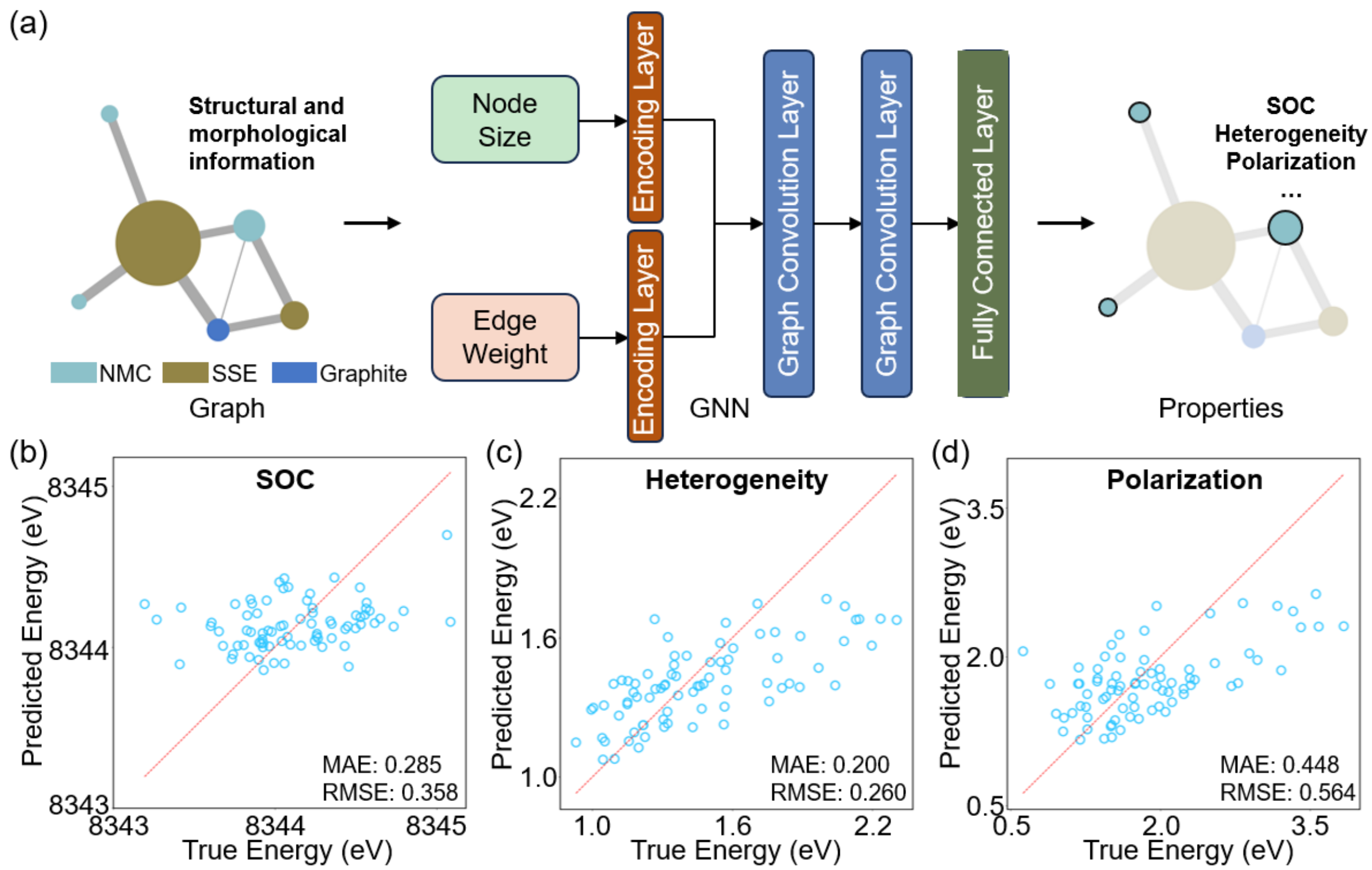

**Figure 6. Predicting local microstructure-property relationship based on graphs. (a)** The demonstration of predicting NMC electrochemical states using GNN, where the input of the GNN is the structural and morphological information of the graph (i.e., node size and edge weight) and the output is the NMC electrochemical state descriptors, (i.e., the intra-particle SOC, heterogeneity, and polarization). **(b)-(d)** GNN prediction versus truth on the testing set for the three electrochemical state descriptors, respectively.

In summary, we developed an ML-enabled workflow for realizing a fast and automated construction of graphs from the experimentally measured X-ray microstructure images of multiphase particulate composites (**Fig. 2**). Using the composite cathode of SSBs as an example, we demonstrated microstructure graphs can enable high-throughput microstructure morphology analysis (**Fig. 3**), the study of the microstructure-property relationships at both the node/edge (**Fig. 4**) and the network (**Fig. 5**) level, and the prediction of the local microstructure-property relationship via GNN (**Fig. 6**). More specifically, based on the automatedly constructed microstructure graphs, we analyzed and directly compared the electrochemical states of the NMC particles with and without the concurrent $Li^+$/electron channels. The results (**Figs**. **4-5**, **S5-8**) provide much-needed microscopic evidence to the importance of TPBs and coupled ionic/electronic channels using ML-enabled graph analysis of the experimentally measured multimodal images. Furthermore, the statistical tests (i.e., Welch's two-sample *t*-test and two-way ANOVA) support our findings (see Supporting Information). This particle-resolved approach provides evidence for the correlation between structural features, i.e., the triple phase boundaries and the concurrent $Li^+$/$e^-$ channels, and the residual electrochemical heterogeneity observed in the relaxed state. These findings suggest that such structural configurations may facilitate electrochemical reactions by mitigating local transport resistances for both $Li^+$ and $e^-$, thereby contributing to a more uniform distribution of overpotential and current density during operations[42, 43]. Thus, our results emphasize not only the importance of selecting appropriate particle morphologies but also the need to control particle self-assembly during electrode fabrication of SSBs. For example, it has been shown that tuning the particle size ratio between the active material and the solid electrolyte can potentially regulate the TPB density and network connectivity[44]. However, establishing more quantitative correlation between the processing parameters and the microstructures of composite cathode represents an open challenge in the field of SSBs. The graph-based microstructure representation demonstrated in this work can potentially function as a critical enabler for the inverse process design, i.e., searching for the process parameters that lead to the desirable microstructures. For example, the graph-theoretic metrics can be used to construct a topology-aware loss function in an active-learning-based search for the process parameters.

We would also like to outline several other challenges and opportunities that are specifically related to solid-state batteries. First, multimodal X-ray imaging in the present work has an ultrahigh spatial resolution of 20 nm, which allows for accurate imaging of the local geometry and topology. Yet, this high resolution comes with the expense of probing a relatively small fraction of the entire composite battery cathode, which may not necessarily be sufficient to serve as a representative volume element (RVE) for the entire cathode. For example, a recent study suggests that a representative volume containing 100 to 1,000 particles corresponds to the scale of heterogeneity in the battery composite electrode[45], which is significantly larger than the number of NMC particles herein. Second, the imaging modality used in the present work, which is different from that in ref. [45]), leads to relatively low image contrast among the background, SSE, and graphite, as shown in Fig. S1. This low contrast makes it exceedingly challenging for human researchers to perform 3D annotation as the ground-truth training data of an ML-based segmentation model. Addressing these two challenges will need more experiments and/or coordinated data sharing as well as the combined use of additional imaging modalities. The present ML-enabled graphic analysis, despite being focused on 2D slices and relatively small microstructure volume, provides a statistically rigorous, self-consistent evaluation of the local microstructure-property relationship at both the individual particle level and the particle network level. Our ML-enabled graph analysis tools can be readily extended to analyze larger-scale (i.e., practically sized) multimodal 3D microstructure datasets once they become available, and the high scalability of the graph-based approach makes it particularly efficient and suitable for harnessing large-scale microstructure data. For example, an earlier study demonstrates a strong scalability of GNN training to thousands of nodes even on a single graphics processing unit (GPU) node[19].

More broadly, our results demonstrate the potential of leveraging graphs to achieve data-driven forward prediction of the microstructure-property relationship and the inverse process design in multiphase particulate composites, where one key challenge has been finding an informative low-dimensional representation of the high-dimensional microstructure data. Specifically, graph-based microstructure representation uniquely allows for encoding both the morphological and topological information of the microstructure as well as phase-dependent local properties (e.g., conductivities) as node/edge features. In the forward prediction, these unique capabilities can potentially lead to an improved prediction accuracy over several baseline ML models, as has previously been demonstrated in polycrystalline microstructures[19]. In the inverse process design, as briefly discussed above, the graph-theoretic metrics can be directly used to construct a topology-aware loss function in active learning that goes beyond the use of simple geometric features such as the mean particle size and size distribution. Overall, we expect the results in this work would stimulate more efforts of constructing and harnessing graph to revolutionize the design and discovery of a wide range of multiphase particulate composite materials.

## Supporting Information

Details of: solid-state battery fabrication and characterization; graph construction enabled by ML-based automated phase segmentation; electrochemical state quantification for NMC particles; graph-theoretic metrics; GNN for predicting NMC electrochemical states; statistical tests for results presenting in Figs. 4e-g and Figs. 5c-e; two-way ANOVA analysis for heterogeneity and polarization; analyses of the potential confounders (particle size, contact area, local porosity, and proximity to boundary layers).

## Acknowledgements

This work is primarily supported by a collaborative National Science Foundation (NSF) Grant under No. CBET-2412157 on the machine learning and computational works (Z. L. and J.-M. H.) and CBET-2412156 on the experimental works (S.D. and Y.L.). Z. L. and J.-M. H. also acknowledge partial support for manuscript preparation from the Air Force Office of Scientific Research under award number FA9550-24-1-0159.

**Supporting Information**

# Machine Learning Enabled Graph Analysis of Particulate Composites: Application to Solid-state Battery Cathodes

Zebin Li[1], Shimao Deng[2], Yijin Liu[2*], Jia-Mian Hu[1*]

[1]*Department of Materials Science and Engineering, University of Wisconsin-Madison, Madison, WI 53706, USA*

[2]*Walker Department of Mechanical Engineering, University of Texas at Austin, Austin, TX 78712, USA*

*E-mails: liuyijin@utexas.edu (Y.L.) or jhu238@wisc.edu (J.-M.H.)

## Solid-state battery fabrication and characterization

**Materials and cell assembly.** $LiNi_{0.8}Co_{0.1}Mn_{0.1}O_2$ (NMC811) powder coated with $LiNbO_3$ (NEI Corporation), along with $Li_6PS_5Cl$ (LPSCl) (NEI Corporation), and lithium metal are employed as the cathode active material, solid-state electrolyte (SSE), and anode, respectively. The composite cathode is formulated by thoroughly mixing NMC811, LPSCl, and graphite in a mass ratio of 6:4:2. For cell fabrication, approximately 150 mg of wet ball-milled LPSCl powder[1] is pressed at 530 MPa for 1 minute using a polyether ether ketone (PEEK) mold to form the SSE layer. Subsequently, around 40 mg of the composite cathode mixture is carefully placed onto the SSE layer and compacted at 867 MPa for 2 minutes. A lithium metal foil, approximately 100 μm thick, is then positioned on the opposite side of the SSE without additional compression. The entire assembly is enclosed in a custom-designed cell casing and tightened to 9 MPa to ensure intimate interfacial contact for electrochemical evaluation.

Galvanostatic charge–discharge testing is conducted using a Land CT2001A battery testing system. The cells are initially cycled at a rate of 0.1 C for the first two formation cycles, after which the current rate is increased to 0.33 C for subsequent cycles. All electrochemical measurements are performed within a voltage range of 2.8 to 4.3 V versus $Li/Li^+$.

**Full-field X-ray imaging.** Three-dimensional full-field transmission X-ray microscopy (TXM) is performed at the full-field X-ray imaging (FXI) beamline (18-ID) of the National Synchrotron Light Source II (NSLS-II), Brookhaven National Laboratory. We clarify that all Ni K-edge TXM measurements were performed ex situ. The time interval between cell preparation and measurement is constrained by beamtime logistics and experimental scheduling, and is typically on the order of 1–2 days. The imaging is carried out with a spatial resolution corresponding to a pixel size of 20 nm. Regarding sample preparation, cathode particles are gently detached from the electrode surface within an argon-filled glove box and immediately loaded into capillary tubes. These are subsequently sealed with epoxy inside the glove box to prevent air exposure. Tomographic projection images are acquired over a 0–180° rotation range under fly scan mode. For the three-dimensional X-ray absorption near-edge structure (XANES) analysis, the incident X-ray energy is scanned across the Ni K-edge, from 8210 eV to 8700 eV, using 63 discrete energy steps. TXM data reconstruction and analysis are performed using the TXM-Wizard software package[2].

### Graph construction enabled by ML-based automated phase segmentation

**Phase segmentation by U-Net.** U-Net is a commonly used architecture build upon convolution neural network for image segmentation proposed in 2015[3]. It is comprised of a symmetric encoder-decoder structure, where the skip connections directly link corresponding layers between the encoder and decoder paths. Such a design of the skip connection enables the U-Net to capture information at different levels, improving its capacity in segmenting complex patterns. Specifically, the encoder is responsible for reducing the spatial dimensions and extracting features (i.e., downsampling) at different scales for the input image, while the decoder then restores the extracted feature map (combined with the features from the corresponding encoder through skip connection) to the original dimension via transposed convolutional operation (i.e., upsampling). The final output of the U-Net is the segmentation map of the input image.

In our work, we develop a customized U-Net enhanced by a dual attention mechanism using PyTorch[4]. The dual attention blocks are incorporated into the decoder skip connections to refine feature fusion by jointly applying class-wise attention, which emphasizes features relevant to semantic class discrimination, and instance-wise attention, which enforces separation between neighboring instances. This design allows the network to suppress irrelevant background activations while enhancing features that contribute to both accurate class assignment and boundary delineation. The overall architecture contains three levels in both the encoder and decoder. Each encoder block consists of two convolutional layers with ReLU activation and dropout, followed by max-pooling for downsampling. The decoder mirrors this structure with transposed convolutions for upsampling, attention-modulated skip connections, and convolutional layers to refine the merged features. The final segmentation map is produced by a $1\times1$ convolutional layer that projects the decoded features onto the target number of classes.

For training, we employ a combined loss function that integrates the Dice loss with a contour-aware binary cross-entropy loss, thereby balancing region consistency with precise boundary localization. Optimization is performed with the AdamW optimizer (learning rate $1\times10^{-4}$, weight decay $1\times10^{-4}$). The dataset (100 images in total) is partitioned into training, validation, and testing with the ratio of 70%, 15%, and 15%. To enhance robustness and generalization, data augmentation is applied to the training set, including random horizontal flips, small rotations, and color jittering. Model performance is evaluated using the mean Intersection-over-Union (mIoU) across all classes (background, SSE, NMC, and graphite), along with class-specific IoU values.

**Post-processing of phase segmentation by watershed algorithm.** The watershed algorithm is a classical image processing technique based on topographic analysis and is particularly effective for separating touching or overlapping objects in images[5]. It treats the image (usually a grayscale image) as a topographic surface, where valleys and peaks represent low-intensity and high-intensity areas, respectively. Then, a flooding process is simulated over the topographic surface (i.e., image), in which water gradually fills the basins from predefined markers. The predefined markers are the starting points of the flooding process, and they correspond to different objects or areas in the image. In particular, a distance map is first calculated, where each pixel within the object is assigned a value representing its distance to the nearest background pixel. The center pixel within the object has the highest distance value (i.e., peak), therefore treated as the predefined marker. As the flooding process from different markers expands, the segmentation map (i.e., boundaries between objects) forms when different basins merge, resulting in the separation of individual objects.

Furthermore, a convexity-based iterative refinement is simultaneously implemented to further enhance the performance of post-processing. Specifically, after each segmentation iteration, the convexity of every separated object (i.e., NMCs) is evaluated. Objects with convexity below a predefined threshold (set to 0.95) are considered potentially under-segmented and are subjected to further watershed processing. Such a strategy allows separating the majority of the connected NMC particles, for example, those shown in **Fig. 2**. For NMC particles with complex, irregular shape or severe agglomeration, occasional over-segmentation or residual merging still occurs. This challenge could be potentially mitigated by adopting more sophisticated learning-based segmentation methods[6] in which case large number of images labeled by domain expert may be needed for the model training.

**Graph construction.** The instance segmentation maps obtained after the post-processing of phase segmentation maps from X-ray images (they are 2D virtual slices, same thereafter) are converted to graph networks, in which the geometric information (e.g., connections among different phases, particle sizes, etc.) of the microstructures for SSBs can be represented in an efficient way. Specifically, each object (e.g., particles of NMC, SSE, and graphite) in the instance segmentation map is considered as a node in the graph, and the areas of the masks for objects are considered as the particles' sizes. Besides, the edges among different nodes represent their adjacency relationships. During segmentation, two objects are typically not allowed to truly overlap (i.e., the same pixel cannot belong to two instances simultaneously). Therefore, the adjacency between two objects is determined using a morphological dilation-based proximity criterion. Specifically, for each object, a binary mask is dilated with a 7×7 square structuring element (i.e., a flat kernel of ones), which expands the object by up to three pixels in all directions under the Chebyshev distance[7]. Two objects are considered adjacent if the dilated mask of one object overlaps with the binary mask of the other. The degree of overlap is quantified as the number of intersecting pixels between the dilated region of the first object and the original region of the second object. This pixel count reflects the extent of near-boundary contact rather than true geometric overlap, since instance masks are mutually exclusive. An undirected edge is added to the graph when this count is nonzero, and the edge weight is defined as a scaled version of this intersection value, thereby encoding the strength of spatial proximity between objects.

**Electrochemical state quantification for NMC particles**

The electrochemical state of NMC particles refers to the Ni oxidation distribution (represented by the Ni K-edge energy distribution) and it is represented by the pixel intensity distribution in the TXM image. By applying the instance segmentation map on the X-ray image, it allows the identification of each NMC particle. Then, the statistics (i.e., mean, standard deviation, and peak distance) of the pixel intensity distribution for each NMC particle are extracted to provide a concise description for its electrochemical state. The pixel distribution is cut at 8340 and 8348 eV according to the Ni adsorption edge. To calculate the peak distance, the pixel histogram of each particle is first smoothed and then analyzed using the *find_peaks* function in Scipy to identify dominant peaks, where a peak is considered dominant if its height is at least 40% of the maximum peak height. If exactly two dominant peaks are present, their separation directly defines the distance; for distributions with multiple peaks, the average positions of the peaks on either side of the reference energy (8344 eV) are compared. It is worth noting that relating the observed Ni K-edge shifts to an approximate change in Li content ($\Delta x$) or state of charge ($\Delta$SOC) would provide a more intuitive sense of the physical and technological significance of the reported differences. However, such quantification would require an appropriate calibration framework, for example using reference spectra with known Ni valence states and linear-combination analysis, which is beyond the scope of the present dataset. The aim of this study is therefore not to assign absolute $\Delta x$ or $\Delta$SOC values from the measured edge shifts, but to use the relative Ni K-edge shifts under identical measurement conditions as an internal indicator of spatial differences in reaction extent.

**Graph-theoretic metrics**

We characterize the connectivity of NMC particles in the graphs using several standard graph-theoretic metrics[8]. The degree of a node quantifies how many neighboring particles it is directly connected to, and the weighted degree extends this by summing the contact strengths along all its edges. The clustering coefficient measures the ratio of the number of triangles in the constructed graph to the total number of geometrically allowable triangles, reflecting how densely interconnected the local environment is. Betweenness centrality evaluates how frequently a node lies on the shortest paths connecting other pairs of nodes, thus indicating whether it acts as a bridge for transport across the network. Closeness centrality measures how close a node is to all other nodes on average, based on shortest-path distances, and captures how efficiently a particle can interact with the entire system. Eigenvector centrality assigns higher importance to nodes that are connected to other well-connected nodes, identifying particles embedded in influential regions of the graph. Moreover, we also count the number of neighbors of each type, i.e., the number of NMC, SSE, and graphite neighbors, around each NMC particle to quantify the local chemical environment.

**GNN for predicting NMC electrochemical states**

Graph neural networks (GNNs) provide a natural framework for analyzing microstructures, as they can capture both node-level attributes (e.g., particle size) and structural information (e.g., inter-particle connections). In our setting, the SSB cathode microstructure is represented as a heterogeneous graph where NMC particles, SSE and graphite phases are nodes of different types, and edges denote physical contacts. By extending message passing to heterogeneous graphs, GNNs are able to learn predictive mappings from microstructural connectivity and attributes to electrochemical states.

We implement a heterogeneous node regression model based on two stacked NNConv layers within a HeteroConv framework, combined with GELU activation, dropout, and LayerNorm, using PyTorch Geometric[9]. The edge attribute (i.e., physical contact strength) is encoded through a small neural network before aggregation, and the residual connection is applied to stabilize training. The target task is to predict NMC electrochemical state descriptors, i.e., intra-particle SOC, heterogeneity, and polarization (they are the means, standard deviations, and peak distances of electrochemical state distributions) at the node level. Data preprocessing, i.e., z-score standardization is performed to improve the training stability. The dataset (73 images in total) is split into training, validation, and testing with the ratio of 60%, 20%, and 20%. The AdamW optimizer is used with the learning rate of $1 \times 10^{-3}$ and the weight decay of $1\times10^{-2}$), gradient clipping (norm $\leq 2$), and an early stopping patience of 50 epochs. The mean squared error (MSE) loss is adopted for the model training. Root mean squared error (RMSE) and mean absolute error (MAE) are used to evaluate the prediction performance.

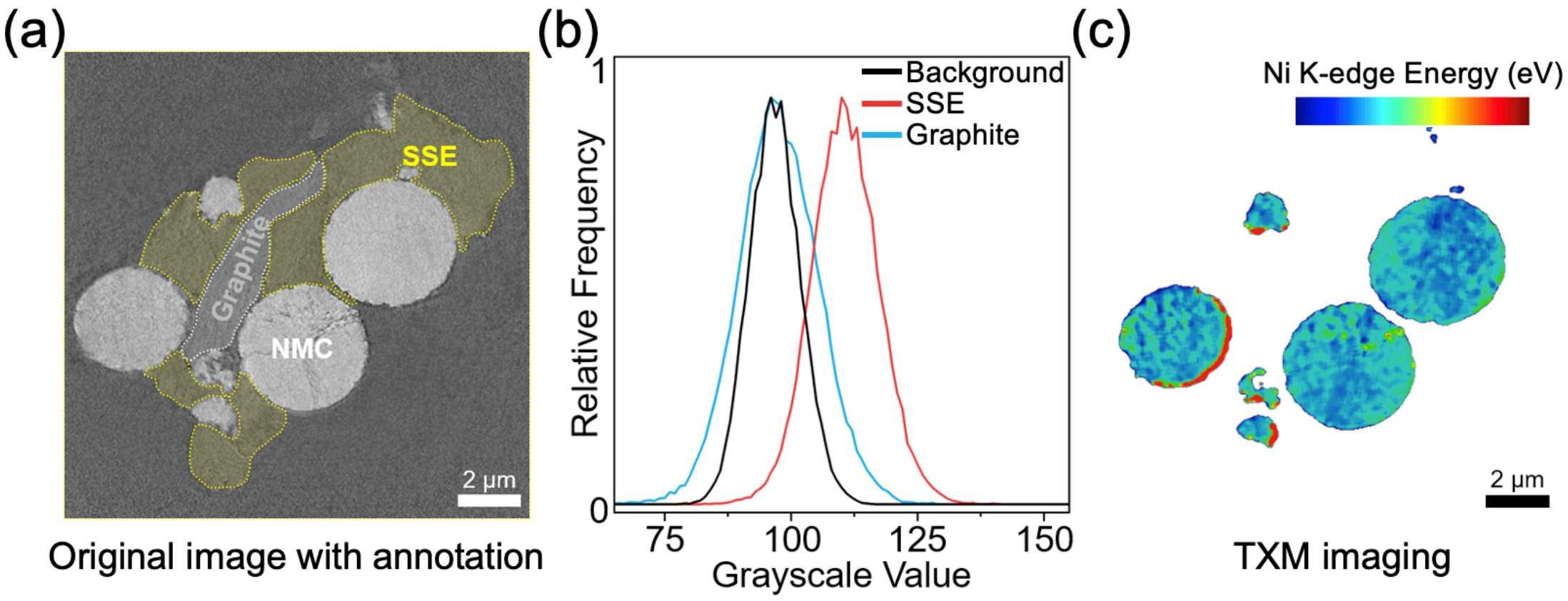

**Figure S1. (a)** Original X-ray image with expert annotation of NMC, SSE, and graphite. **(b)** The pixel intensity distributions of the SSE, the graphite, and the background in (a), reveal their similarity. **(c)** Example of local electrochemical states (i.e., Ni oxidation states represented by the Ni K-edge energy) of NMC particles obtained by TXM imaging for (a).

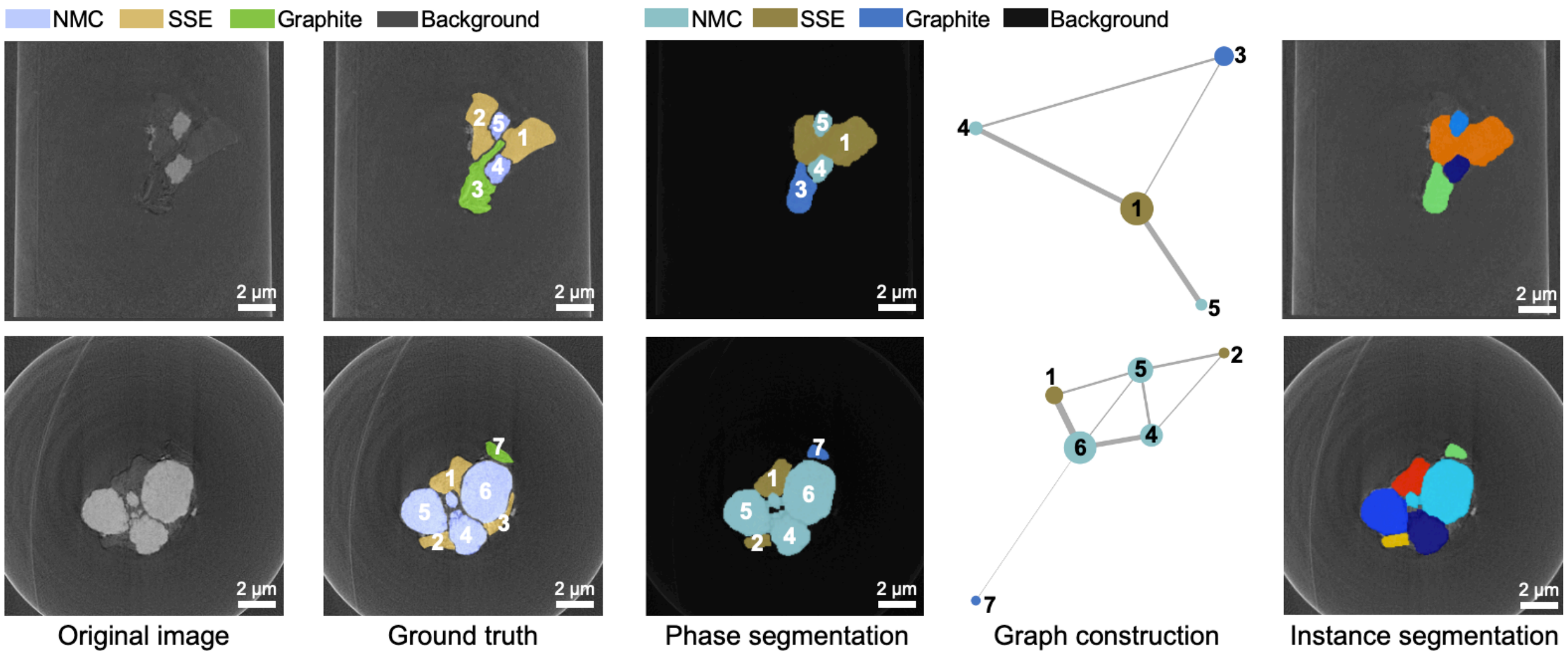

**Figure S2.** Example results of graph construction enabled by ML-based automated phase segmentation.

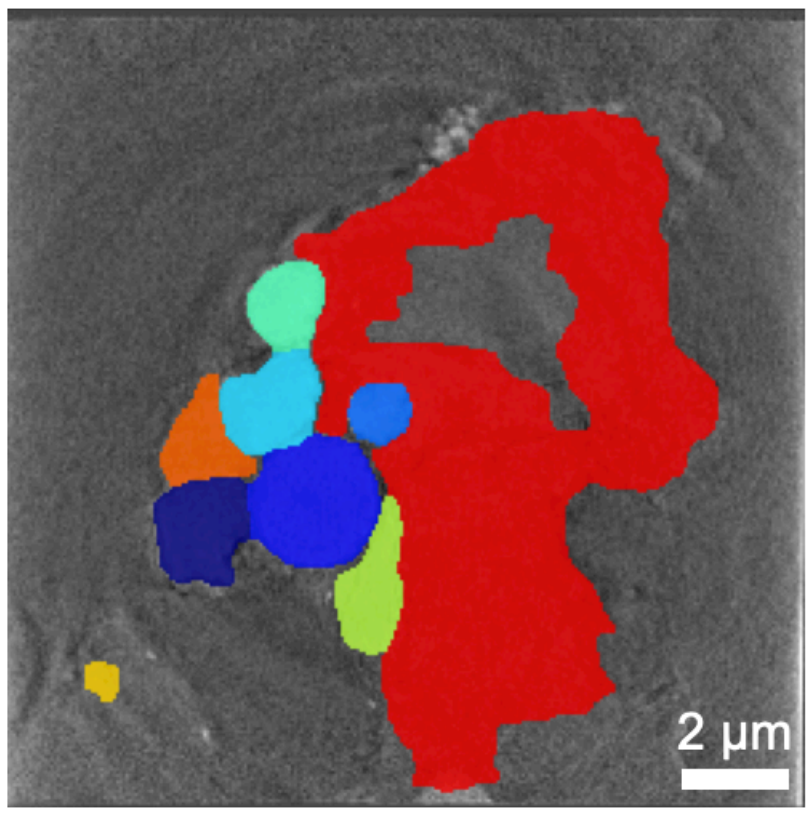

**Figure S3.** Example of the instance segmentation result after performing the watershed algorithm. Each object in the X-ray image is represented by a unique color.

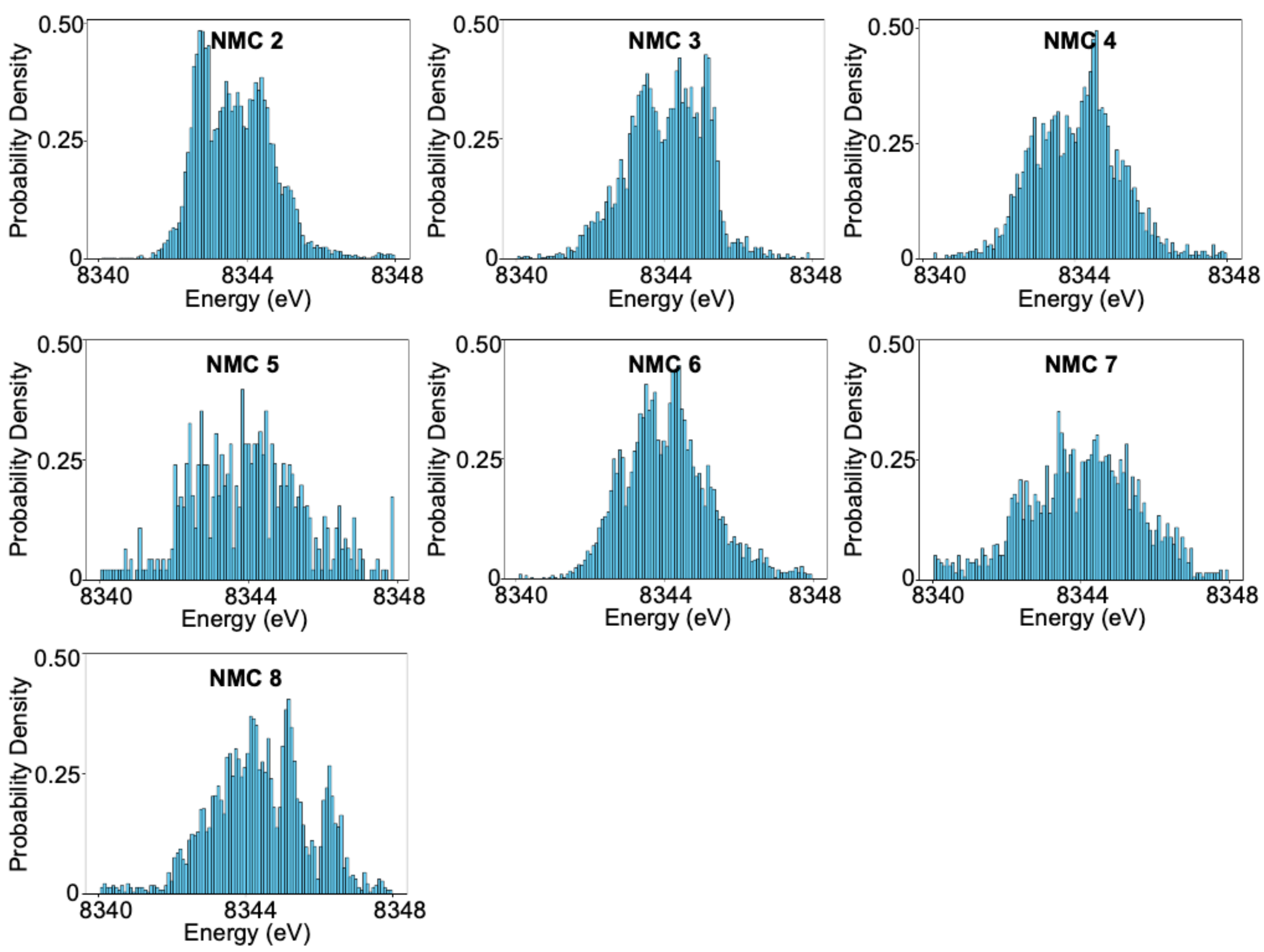

**Figure S4.** The electrochemical state (i.e., Ni oxidation state, represented by the Ni K-edge energy) distributions of NMC particles #2 – #8 in **Fig. 4a**.

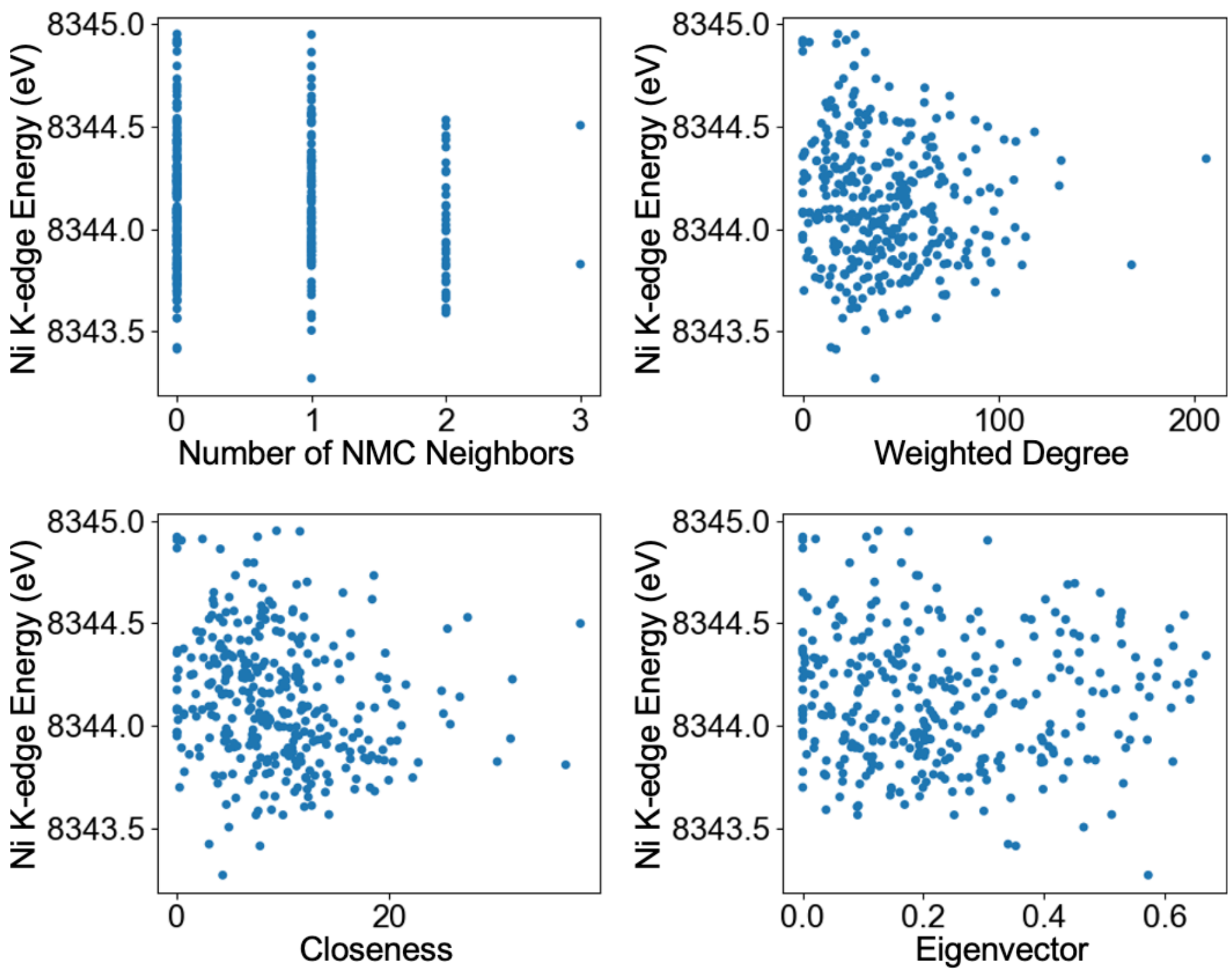

**Figure S5.** Scatter plots of the top four graph-theoretic metrics with the strongest absolute Pearson correlations with SOC.

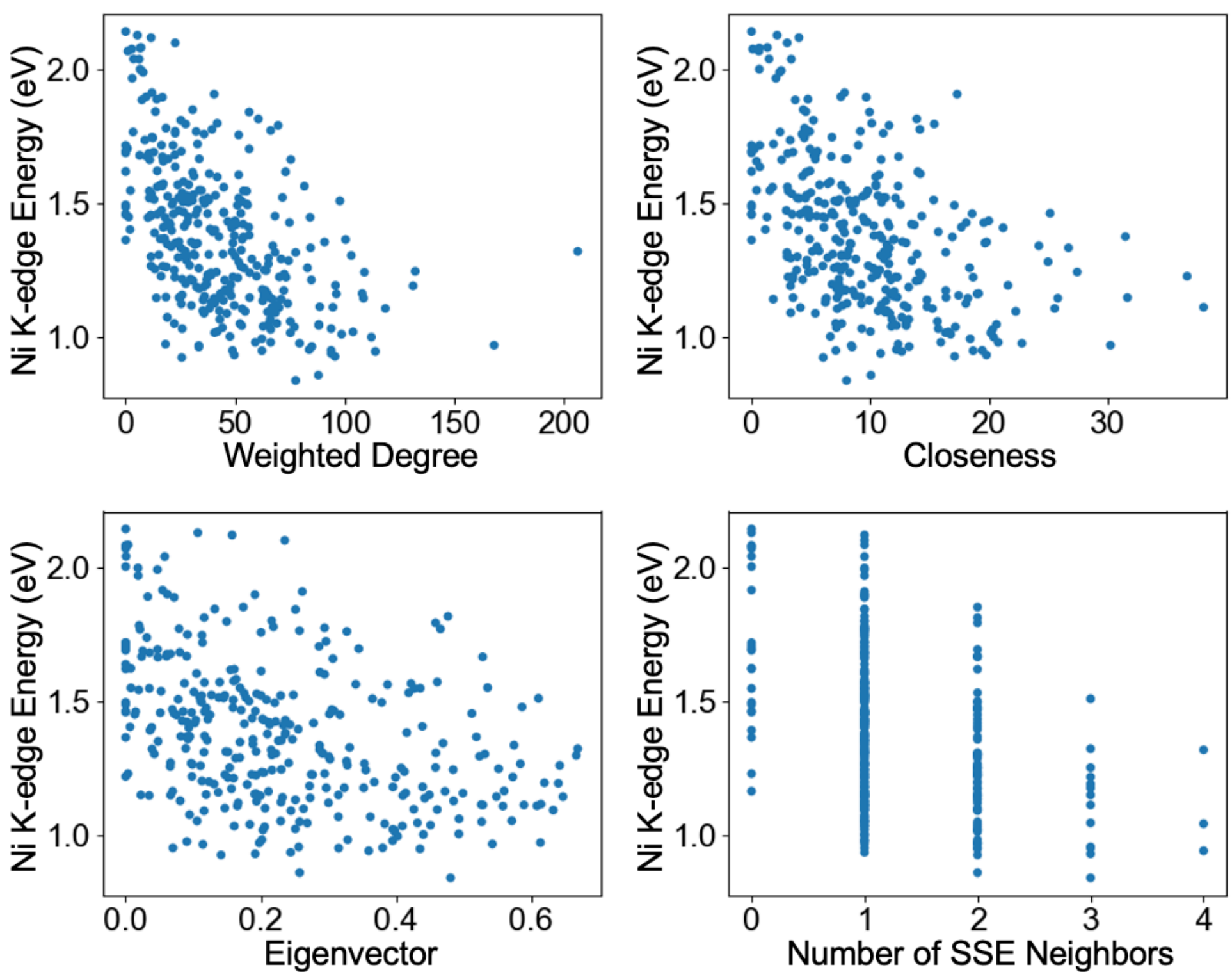

**Figure S6.** Scatter plots of the top four graph-theoretic metrics with the strongest absolute Pearson correlations with heterogeneity.

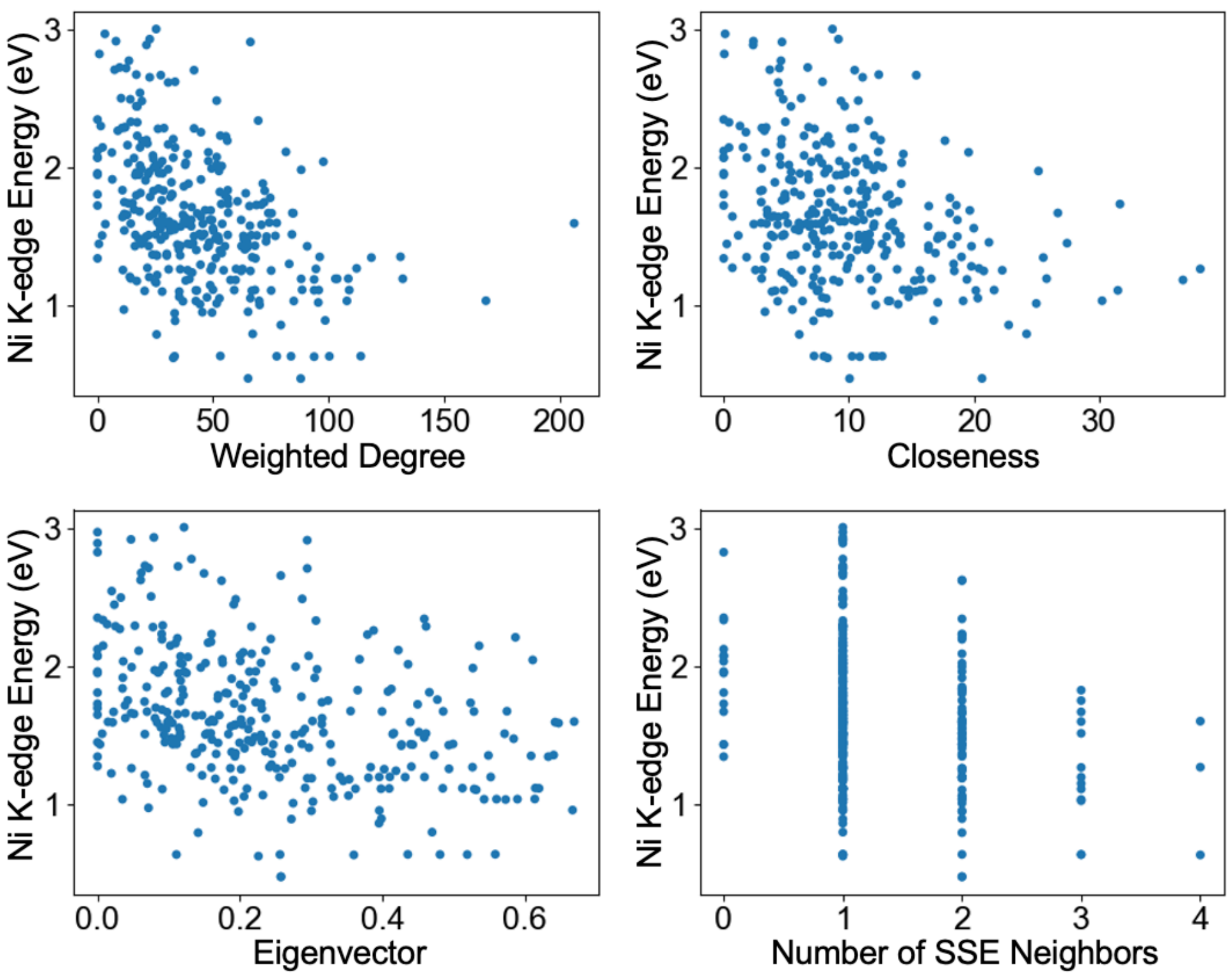

**Figure S7.** Scatter plots of the top four graph-theoretic metrics with the strongest absolute Pearson correlations with polarization.

**Statistical tests for results presenting in Figs. 4e-g and Figs. 5c-e**

We employ Welch's two-sample *t*-test with a significance level of $\alpha = 0.05$ on the data presented in Figs. 4e-g and Figs. 5c-e, since it does not assume equal variances and provides a robust inference under unequal sample sizes, which fits our cases[10]. The results are shown in Table S1. The effect sizes are evaluated by Cohen's *d* measurement[11]. From Table S1, it can be observed that the statistical tests support our conclusion that the TPB and the concurrent existence of $Li^+$ and $e^-$ pathways have positive impact on the heterogeneity (Fig. 4f and Fig. 5d) and polarization (Fig. 4g and Fig. 5e). Although the results for SOC (Fig. 4e and Fig. 5c) are not statistically significant, they are consistent with our findings in the manuscript that the averaged SOCs of the NMC particles remain largely independent of their involvement in a TPB or concurrent existence of $Li^+$ and $e^-$ pathways.

Table S1. Statistical Test Results for Figs. 4e-g and Figs. 5c-e

| Figures | P-Values | Effect Sizes | Confidence Intervals |
|---|---|---|---|
| Fig.4e | 0.542 | 0.08 | [-0.075, 0.144] |
| Fig.4f | 0.00165 | -0.43 | [-0.212, -0.053] |
| Fig.4g | 0.028 | -0.29 | [-0.377, -0.025] |
| Fig.5c | 0.346 | -0.10 | [-0.125, 0.044] |
| Fig.5d | $2.25\times10^{-8}$ | -0.58 | [-0.233, -0.114] |
| Fig.5e | $1.17\times10^{-7}$ | -0.53 | [-0.486, -0.228] |

**Two-way ANOVA analysis for heterogeneity and polarization**

We further conduct a two-way Analysis of Variance (ANOVA) to investigate the effects of TPB involvement (present vs absent) and concurrent $Li^+$ and $e^-$ transport pathways (present vs absent) on heterogeneity and polarization. We analyze heterogeneity and polarization separately and do not investigate SOC due to its independence of TPB and concurrent $Li^+/e^-$ pathways involvement. To address the non-normality and heteroscedasticity commonly observed in electrochemical characterization data, we apply a log-transformation to the heterogeneity and the polarization data (i.e., the standard deviation and the peak distance of the electrochemical state represented by the Ni K-edge energy distribution). The two-way ANOVA results for polarization and heterogeneity are shown in Table S2 and S3.

Table S2. Two-way ANOVA Results for Heterogeneity

| | Sum of Squares | Degrees of freedom | F-statistic | P-value |
|---|---|---|---|---|
| With TPB | 0.073097 | 1 | 5.101139 | 0.024532 |
| With concurrent $Li^+/e^-$ pathways | 0.347206 | 1 | 24.230103 | 0.000001 |
| Interaction of TPB/concurrent pathways | 0.001998 | 1 | 0.139463 | 0.709044 |
| Residual | 4.958020 | 346 | NaN | NaN |

Table S3. Two-way ANOVA Results for Polarization

| | Sum of Squares | Degrees of freedom | F-statistic | P-value |
|---|---|---|---|---|
| With TPB | 0.096885 | 1 | 1.923031 | 0.166416 |
| With concurrent $Li^+/e^-$ pathways | 0.979693 | 1 | 19.445506 | 0.000014 |
| Interaction of TPB/concurrent pathways | 0.003176 | 1 | 0.063037 | 0.801908 |
| Residual | 17.431979 | 346 | NaN | NaN |

These results suggest that TPB involvement and concurrent $Li^+/e^-$ pathways have distinct contributions to electrochemical behavior. After accounting for concurrent pathway accessibility, TPB involvement remains significantly associated with heterogeneity. While for polarization, only concurrent pathway remains significant, the TPB involvement does not show an independent effect. The two-way ANOVA results are mostly consistent with our previous statistical analysis (Table S1 in the Supporting Information). Specifically, TPB involvement is significantly associated with heterogeneity (Fig. 4f), while concurrent pathway accessibility is significantly associated with both heterogeneity and polarization (Fig. 5d–e). For polarization, the marginal significance observed for TPB involvement in the pairwise test ($p = 0.028$ in Fig. 4g) does not persist after accounting for concurrent pathway accessibility in the two-way ANOVA, indicating that this effect is not independent.

These results suggest that concurrent $Li^+/e^-$ pathway accessibility has a stronger and more consistent effect than TPB involvement across both heterogeneity and polarization. Nevertheless, TPB involvement still plays an important role, particularly as a local structural feature influencing heterogeneity. In addition, we would like to emphasize that the analyzed experimental data are noisy and 2D, which may introduce bias in the quantitative estimates of the effects. The primary aim of the analyses shown in Figs. 4 and 5 is to demonstrate the use of graph representation for downstream tasks, thereby corroborating the effectiveness of graph representation in revealing the local microstructure-property relationship in multiphase particulate composites.

**Analyses of the potential confounders (particle size, contact area, local porosity, and proximity to boundary layers)**

**Particle size.** To decouple NMC particle size from TPB abundance, we investigate the correlation between the particle size and the electrochemical state (i.e., SOC, heterogeneity, and polarization) for NMCs involved in one TPB and two TPBs, since most of NMCs involved in TPBs fall into these two categories according to Fig. 3b. The scatter plots are shown in Figure S8. The first row shows results for NMCs involved in one TPB, while the second row for those involved in two TPBs. It can be observed that there is no clear pattern, which indicates that particle size alone does not have significant impact on the electrochemical state, suggesting that the TPB abundance plays a more critical role.

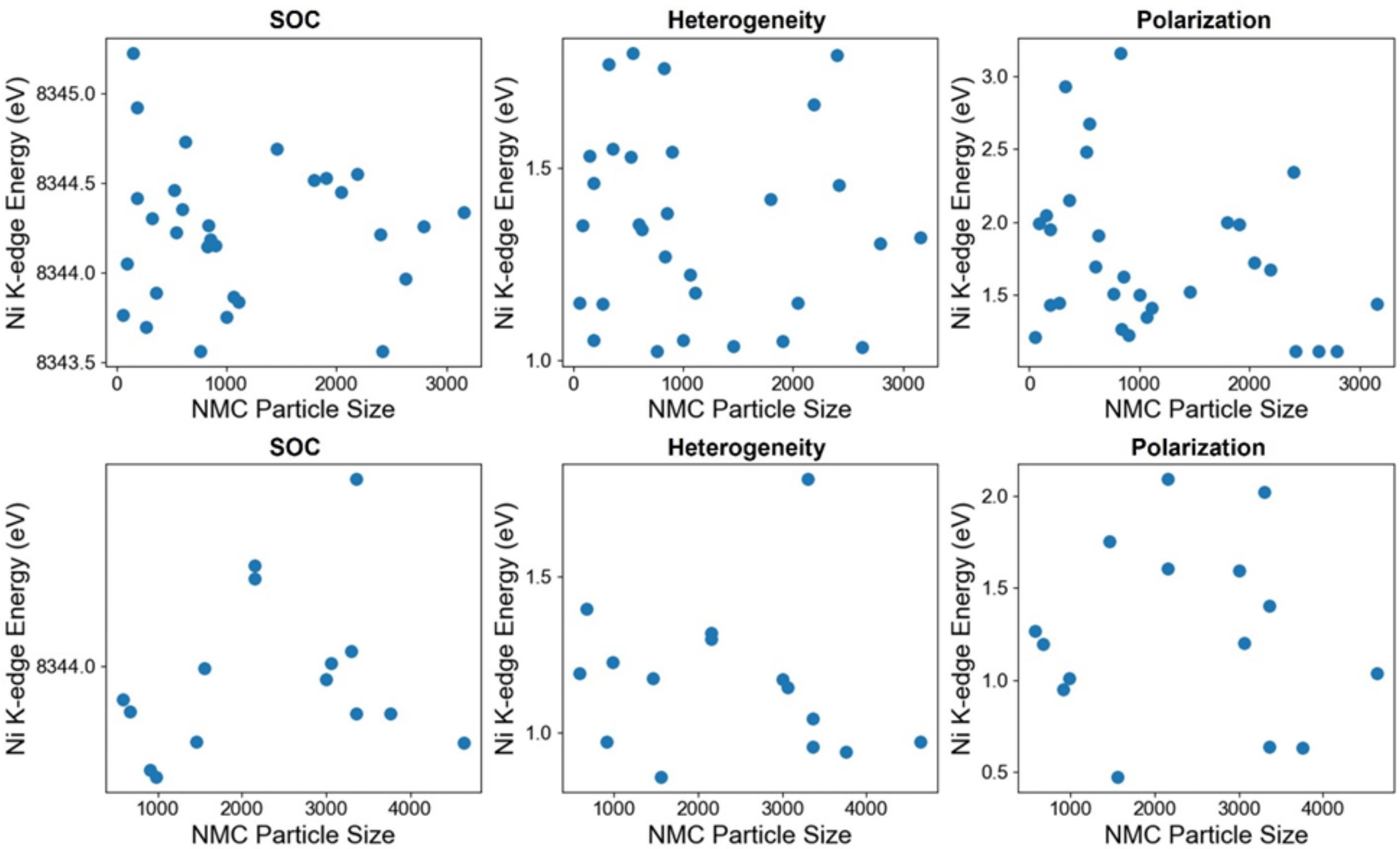

**Figure S8.** Scatter plots of the electrochemical state (i.e., SOC, heterogeneity, and polarization) with respect to the NMC particles size for NMCs involved in one TPB (the first row) and two TPBs (the second row).

**Contact area and local porosity.** Contact area and porosity are structurally coupled with TPB formation and ion/electron transport pathways. In our analysis, we explicitly quantify ion and electron channel connectivity at the particle surface, which inherently captures the effect of contact area and local porosity. Therefore, these factors are not independent variables but are embedded within the TPB/channel descriptors used in this study.

**Proximity to boundary layers.** Proximity to interfaces may influence ionic and electronic conductivity, particularly ionic transport. However, since our analysis is performed on a particle-resolved basis within the same electrode environment, particles with and without TPBs are subject to comparable boundary conditions. Thus, while boundary effects may contribute to local variations, they do not systematically bias the TPB-based comparisons.

In summary, though these structural parameters may contribute to local heterogeneity, TPB abundance captures the dominant structural contribution among these correlated descriptors.

**Supporting References**